\documentstyle[aps,epsf]{revtex}
\begin{document}
\title{
{\it Submitted to Phys. Rev. A. 3$^{rd}$ July, 1996 \vskip 3mm}
Separation of the Exchange-Correlation Potential into Exchange
plus Correlation: \\ an Optimized Effective Potential Approach}

\author{Claudia Filippi}
\address{Laboratory of Atomic and Solid State Physics and Cornell Theory
Center, Cornell University, Ithaca, New York 14853}
\author{C. J. Umrigar}
\address{Cornell Theory Center and Laboratory of Atomic and Solid State
Physics, Cornell University, Ithaca, New York 14853}
\author{Xavier Gonze}
\address{Unit\'e P.C.P.M., Universit\'e Catholique de Louvain,
B-1348 Louvain-la-Neuve, Belgium}
\maketitle
\begin{abstract}
Most approximate exchange-correlation functionals used within density 
functional theory are constructed as the sum of two distinct 
contributions for exchange and correlation. Separating the exchange component 
from the entire functional is useful since, for exchange, exact relations 
exist under uniform density scaling and spin scaling. 
In the past, accurate exchange-correlation potentials have been 
generated from essentially exact densities
but they have not been correctly decomposed into their separate 
exchange and correlation components (except for two-electron systems).
Using a recently proposed method, equivalent to the solution of an optimized 
effective potential problem with the corresponding orbitals replaced by the 
exact Kohn-Sham orbitals, we obtain the separation according to the density 
functional theory definition.
We compare the results for the Ne and Be atoms with those obtained by the 
previously used approximate separation scheme.
\end{abstract}

\bigskip

\section{Introduction}
\label{s0}

Within density functional theory (DFT), the ground state energy of an
interacting system of electrons in an external potential can be written as
a functional of the ground state electronic density~\cite{HK}. 
In the Kohn-Sham formulation of density functional theory~\cite{KS}, the
ground state density is obtained as the density of a system of non-interacting 
electrons in an effective local potential.
Although density functional theory is in principle exact, the energy functional 
contains an unknown quantity, called the exchange-correlation energy,
$E_{\rm xc}\left[\rho\right]$.
The effective potential for the fictitious non-interacting system is the sum
of the external potential, the Hartree potential and the exchange-correlation
potential, which is the functional derivative 
with respect to the density of $E_{\rm xc}\left[\rho\right]$.
The density functional theory definition of the separate exchange and 
correlation components of $E_{\rm xc}\left[\rho\right]$ is based on the 
non-interacting system and is such that the resulting exchange functional 
has properties that are useful guides in the construction of an approximate 
exchange. 
Consequently, most approximate exchange-correlation functionals are also 
constructed as the sum of two distinct contributions for exchange and 
correlation. 

In the past, exchange-correlation potentials and energies, of varying 
degrees of accuracy, have been determined by generating a density for the 
system of interest and then computing an exchange-correlation potential 
that yields the desired density as the ground state solution for the 
fictitious non-interacting system.
In this context, researchers have used charge densities calculated by 
quantum chemistry methods for atoms~\cite{S79,vonBarth84,AP,AS,NM,D90,%
Chen94,LB1,MZ,Ludena95,he} and molecules~\cite{BBS,GLB,IH}, as well as 
Quantum Monte Carlo methods for atoms~\cite{proc,UG} and for a model 
semiconductor~\cite{KG}.
The subsequent inverse problem, namely the search of the corresponding
exchange-correlation potential, has been performed using a variety of 
different techniques. For example, the exchange-correlation potential has been
determined by expanding it in a set of basis functions and varying 
the expansion coefficients to reproduce the accurate density~\cite{AP,UG}.

With the exception of two-electron systems, these accurate exchange-correlation 
potentials have never been separated into their exchange and correlation 
components according to the density functional theory definition. 
For many-electron systems, an approximate scheme was used where the exchange 
potential was defined as the potential yielding the Hartree-Fock density and 
the correlation potential as the difference of the accurate 
exchange-correlation potential and this approximate exchange potential. 
In this paper, following the approach proposed by G\"orling and Levy~\cite{GL},
we obtain for the first time the correct separation of accurate 
exchange-correlation potentials for the Be atom and the Ne atom.

In Sec.~\ref{s1}, we briefly introduce density functional theory and its 
Kohn-Sham formulation. In Sec.~\ref{s2}, we derive the formulae used to
determine the decomposition of the exchange-correlation potential into exchange
and correlation. A comparison with approximate separation schemes is given 
in Sec.~\ref{s3}. In Appendix~\ref{a1}, we describe the method for the 
special case of closed shell systems.

\section{Theoretical background}
\label{s1}

Density functional theory provides an expression of the ground
state energy of a system of interacting electrons in an external potential
as a functional of the ground state electronic density \cite{HK}. 
Let us assume for simplicity that the spin polarization of the system
of interest is identically zero.
In the Kohn-Sham formulation of density functional theory \cite{KS}, the ground 
state density is written in terms of single-particle orbitals obeying the
equations in atomic units ($\hbar=e=m=1$):
\begin{eqnarray}
\left\{-\frac{1}{2}\nabla^2+v_{\rm ext}({\bf r})+\int \frac{\rho({\bf r}')}
{\left|{\bf r}-{\bf r}'\right|}{\rm d}{\bf r}'+v_{\rm xc}\left(
\left[\rho\right];{\bf r}\right)\right\}\psi_i=\epsilon_i\psi_i,
\label{KS}
\end{eqnarray}
where
\begin{eqnarray}
\rho({\bf r})=\sum_{i=1}^N\left|\psi_i({\bf r})\right|^2.
\label{rho}
\end{eqnarray}
The electronic density is constructed by summing over the {\it N} lowest
energy orbitals where {\it N} is the number of electrons.
$v_{\rm ext}({\bf r})$ is the external potential.
The exchange-correlation potential $v_{\rm xc}\left(\left[\rho\right];
{\bf r}\right)$ is the functional derivative of the exchange-correlation
energy $E_{\rm xc}\left[\rho\right]$ that enters in the expression for
the total energy of the system:
\begin{eqnarray}
E=-\frac{1}{2}
\sum_{i=1}^N\int\psi_i\nabla^2\psi_i\,{\rm d}{\bf r}
+\int\,\rho\left({\bf r}\right)v_{\rm ext}
\left({\bf r}\right)\,{\rm d}{\bf r}
+\frac{1}{2}\int\!\!\int\frac{\rho({\bf r})
\rho({\bf r}')}{\left|{\bf r}-{\bf r}'\right|} {\rm d}{\bf r}\,{\rm d}{\bf r}'
+E_{\rm xc}\left[\rho\right].
\label{eq0}
\end{eqnarray}
The exchange-correlation functional is written as the sum of two separate
contributions for exchange and correlation,
\begin{eqnarray}
E_{\rm xc}\left[\rho\right]=E_{\rm x}\left[\rho\right]+
E_{\rm c}\left[\rho\right].
\end{eqnarray}
The definition of the exchange energy is in terms of the non-interacting
wave function $\Phi_0$, the Slater determinant constructed from the Kohn-Sham
orbitals, as
\begin{eqnarray}
{\rm E}_{\rm x}\left[\rho\right]=
\left<\Phi_0\right|V_{\rm ee}\left|\,\Phi_0\right>-
\frac{1}{2}\int\!\!\int\frac{\rho({\bf r})
\rho({\bf r}')}{\left|{\bf r}-{\bf r}'\right|}
{\rm d}{\bf r}\,{\rm d}{\bf r}',\label{enx0}
\end{eqnarray}
where $V_{\rm ee}$ is the electron-electron interaction.
This definition differs from the conventional quantum chemistry definition 
of $E_{\rm x}$ as the exchange energy in a Hartree-Fock calculation, given
by the same expression as in Eq.~\ref{enx0} but with the Kohn-Sham
determinant replaced by the Hartree-Fock determinant.
The separation of the exchange-correlation functional into exchange and
correlation yields a corresponding splitting of the exchange-correlation
potential into $v_{\rm x}\left(\left[\rho\right];{\bf r}\right)$ and
$v_{\rm c}\left(\left[\rho\right];{\bf r}\right)$.
In this formulation, the essential unknown quantity is the exchange-correlation
energy $E_{\rm xc}\left[\rho\right]$. If the functional form of $E_{\rm xc}
\left[\rho\right]$, and consequently the exchange-cor\-re\-la\-tion potential,
were available, we could solve the {\it N}-electron problem by finding the
solution of a set of single-particle equations.

The exchange functional, as defined in Eq.~\ref{enx0}, scales under uniform 
density scaling, 
$\rho_\lambda({\bf r})=\lambda^3\rho(\lambda{\bf r})$,~\cite{LP} as
\begin{eqnarray}
E_{\rm x}\left[\rho_\lambda\right]=\lambda E_{\rm x}\left[\rho\right],
\end{eqnarray}
and its spin polarized version is simply given in terms of the unpolarized 
exchange functional~\cite{OP} as 
\begin{eqnarray}
E_{\rm x}\left[\rho_\uparrow,\rho_\downarrow\right]=\frac{1}{2}\left\{
E_{\rm x}\left[2\rho_\uparrow\right]+E_{\rm x}\left[2\rho_\downarrow\right]
\right\}.
\end{eqnarray}
Clearly, the separation of the exchange-correlation functional according to 
Eq.~\ref{enx0} is useful since, for exchange, only an approximation for the 
unpolarized functional needs to be sought and the behavior under uniform 
scaling determines how derivatives of the density combine with the density in 
an approximate exchange functional:
\begin{eqnarray}
E^{\rm approx}_{\rm x}\left[\rho\right]=\int\,\rho({\bf r})^{4/3}\,
F(\,\left|\nabla\rho({\bf r})\right|/\rho({\bf r})^{4/3},
\nabla^2\rho({\bf r})/\rho({\bf r})^{5/3},\ldots\,)\,{\rm d}{\bf r}.
\end{eqnarray}

\section{Separation of $v_{\rm xc}$ into $v_{\rm x}$ plus $v_{\rm c}$}
\label{s2}

For the special case of two electrons in a singlet state, the separation of the 
exchange-correlation potential into exchange plus correlation is quite simple 
since the exchange potential is simply given by the condition that it
cancels the self-interaction term in the Hartree potential.
On the other hand, for many-electron systems, this decomposition into exchange 
and correlation components has never been done.
In previous work~\cite{AP,AS,Chen94,proc,Ludena95}, the
exchange potential was defined as the difference of the effective Kohn-Sham
potential yielding the Hartree-Fock density and the sum of the Hartree
and the external potentials.
The correlation potential was then obtained as the difference of the
exchange-correlation potential corresponding to the exact density and the above 
approximate exchange potential. Note that this ``exchange'' potential is not an
exchange-correlation potential since we are subtracting the wrong external
potential: the Hartree-Fock density is the true ground state density
for a Hamiltonian with an external potential different than the original one.
However, it is also not the exchange potential corresponding to the
Hartree-Fock density (although very close to it) since it is not the
functional derivative with respect to the density of the exchange energy
evaluated for the orbitals obtained from the effective potential yielding
the Hartree-Fock density.
Therefore, this separation scheme is incorrect: it involves two densities,
the exact and the Hartre-Fock densities, and, moreover, the potential used
for exchange is only approximately equal to the exchange potential
corresponding to the Hartree-Fock density.

We follow G\"orling and Levy \cite{GL} in showing how to
separate the exchange-correlation potential into exchange and correlation.
We consider a spin unpolarized system.
If we assume that the density $\rho$ is non-interacting {\it v}-representable,
it can be expressed as in Eq.~\ref{rho} in terms of single-particle orbitals
$\{\psi_i\}$ of the Kohn-Sham potential $v_{\rm s}\left({\bf r}\right)$,
\begin{eqnarray}
v_{\rm s}\left({\bf r}\right)=v_{\rm ext}\left({\bf r}\right)+
\int\frac{\rho\left({\bf r}'\right)}{\left|{\bf r}-{\bf r}'\right|}
{\rm d}{\bf r}'+v_{\rm xc}\left(\left[\rho\right];{\bf r}\right).
\label{vKS}
\end{eqnarray}
The exchange energy is a functional of the density but can also be expressed in
terms of the Kohn-Sham orbitals $\{\psi_i\}$ (Eq. \ref{enx0}) as
\begin{eqnarray}
{\rm E}_{\rm x}\left[\rho\right]=
-\frac{1}{2}\sum_{i=1}^N\sum_{j=1}^N\delta_{m_{s_i},m_{s_j}}\int\int
\frac{\psi_i^*({\bf r})\psi_j^*({\bf r}')\psi_j({\bf r})\psi_i({\bf r}')}
{\left|{\bf r}-{\bf r}'\right|}\,{\rm d}{\bf r}\,{\rm d}{\bf r}',
\label{enx1}
\end{eqnarray}
where the $\delta$-function is over the spin quantum numbers of the $i$-th
and $j$-th orbitals.
We evaluate the functional derivative of the exchange energy functional
with respect to the Kohn-Sham potential as
\begin{eqnarray}
\frac{\delta {\rm E}_{\rm x}\left[\rho\right]}{\delta v_{\rm s}({\bf r})}=
\int\frac{\delta {\rm E}_{\rm x}\left[\rho\right]}{\delta
\rho({\bf r}')}\frac{\delta \rho({\bf r}')}{\delta v_{\rm s}({\bf r})}\,
{\rm d}{\bf r}'
=\int v_{\rm x}\left(\left[\rho\right];{\bf r}'\right)
\sum_{i=1}^N\left(\psi_i^*({\bf r}')\frac{\delta \psi_i({\bf r}')}
{\delta v_{\rm s}({\bf r})}+\frac{\delta \psi_i^*({\bf r}')}
{\delta v_{\rm s}({\bf r})}\psi_i({\bf r}')\right)\,
{\rm d}{\bf r}'.
\label{oep1}
\end{eqnarray}
On the other hand, since the exchange functional can be written as a function
of the orbitals (Eq.~\ref{enx1}), we also have
\begin{eqnarray}
\frac{\delta {\rm E}_{\rm x}\left[\rho\right]}{\delta v_{\rm s}({\bf r})}=
\sum_{i=1}^N\int\left(\frac{\delta {\rm E}_{\rm x}\left[\rho\right]}
{\delta \psi_i({\bf r}')}\frac{\delta \psi_i({\bf r}')}
{\delta v_{\rm s}({\bf r})}
+\frac{\delta {\rm E}_{\rm x}\left[\rho\right]}{\delta \psi_i^*({\bf r}')}
\frac{\delta \psi_i^*({\bf r}')}{\delta v_{\rm s}({\bf r})}\right)\,
{\rm d}{\bf r}'.
\label{oep2}
\end{eqnarray}
If we combine Eqs.~\ref{oep1} and \ref{oep2}, we obtain the integral equation
\begin{eqnarray}
\int v_{\rm x}\left(\left[\rho\right];{\bf r}'\right)
{\cal K}\left({\bf r}',{\bf r}\right){\rm d}{\bf r}'=
{\cal Q}\left({\bf r}\right),
\label{vxeq}
\end{eqnarray}
where the kernel ${\cal K}\left({\bf r}',{\bf r}\right)$ and the right hand
side ${\cal Q}\left({\bf r}\right)$ depend on the orbital $\{\psi_i\}$ and
their functional derivative with respect to the potential $v_{\rm s}\left(
{\bf r}\right)$.
This integral equation is equivalent to the one solved in the optimized
effective potential method (OEP) where the KS orbitals are replaced by the 
OEP orbitals~\cite{OEP}.
The functional derivatives of the orbitals $\delta \psi_i({\bf r})/
\delta v_{\rm s}({\bf r}')$ can be expressed in terms of the Green's function
$G_i\left({\bf r},{\bf r}'\right)$ as
\begin{eqnarray}
\frac{\delta \psi_i({\bf r})}{\delta v_{\rm s}({\bf r}')}
=-G_i\left({\bf r},{\bf r}'\right)\psi_i({\bf r}'),
\end{eqnarray}
where $G_i\left({\bf r},{\bf r}'\right)$ satisfies the differential equation
\begin{eqnarray}
\left(-\frac{1}{2}\nabla^2+v_{\rm s}({\bf r})-\epsilon_i\right)
G_i\left({\bf r},{\bf r}'\right)
=\delta\left({\bf r}-{\bf r}'\right)-\psi_i({\bf r})\psi_i^*({\bf r}').
\label{g.gen}
\end{eqnarray}
By knowing the exchange-correlation potential, the KS orbitals and eigenvalues,
we can compute the Green's functions $\{G_i\}$ and, consequently, the kernel
$\cal K$ and the function $\cal Q$.
If we express the exchange potential as a linear combination of basis 
functions, Eq.~\ref{vxeq} can be rewritten as a non-homogeneous set of
linear equations for the coefficients of the expansion of the potential 
in the basis set.

Once the exchange potential is determined, the correlation potential is simply 
obtained as the difference:
\begin{eqnarray}
v_{\rm c}\left([\rho];{\bf r}\right)=v_{\rm xc}\left([\rho];{\bf r}\right)
-v_{\rm x}\left([\rho];{\bf r}\right).\label{vcorr}
\end{eqnarray}

In Appendix~\ref{a1}, the equations derived in this section are rewritten
for the case of a closed shell system.

\section{Comparison with approximate separation schemes}
\label{s3}

For the Be atom and the Ne atom, we calculate the exchange potentials as
explained in the previous section. The correlation potentials are determined
as the difference of the accurate exchange-correlation potentials and the
exchange components (Eq.~\ref{vcorr}). We already mentioned that, in the past, 
an approximate ``exchange'' potential was instead used, given by the effective 
potential yielding the Hartree-Fock density minus the Hartree and the external 
potentials:
\begin{eqnarray}
\tilde{v}_{\rm x}([\rho_{\rm HF}];{\bf r})=v_s([\rho_{\rm HF}];{\bf r})-
\int{\rm d}{\bf r}'\frac{\rho_{\rm HF}({\bf r'})}
{\left|{\bf r}-{\bf r}'\right|}-v_{\rm ext}({\bf r}),
\end{eqnarray}
where we introduced an explicit dependence of the Kohn-Sham potential $v_s$
on the density reproduced by $v_s$.
The potential $\tilde{v}_{\rm x}$ is very close to the exchange potential 
corresponding to the Hartree-Fock density, 
$v_{\rm x}([\rho_{\rm HF}];{\bf r})$.
To determine $v_{\rm x}([\rho_{\rm HF}];{\bf r})$, we can use the same scheme 
explained in the previous section with the orbitals given by the Kohn-Sham 
orbitals corresponding to the effective potential yielding the Hartree-Fock 
density instead of the exact density. 
We denote by $v_{\rm c}^{\rm A}$ and $v_{\rm c}^{\rm B}$ the correlation 
potentials determined as the difference of the accurate $v_{\rm xc}$ and
$\tilde{v}_{\rm x}([\rho_{\rm HF}];{\bf r})$ and 
$v_{\rm x}([\rho_{\rm HF}];{\bf r})$ respectively:
\begin{eqnarray}
v_{\rm c}^{\rm A}({\bf r})=v_{\rm xc}([\rho];{\bf r})-
\tilde{v}_{\rm x}([\rho_{\rm HF}];{\bf r}),\;\;\;\;
v_{\rm c}^{\rm B}({\bf r})=v_{\rm xc}([\rho];{\bf r})-
v_{\rm x}([\rho_{\rm HF}];{\bf r}).
\label{vcorrapprox}
\end{eqnarray}

As discussed in Ref.~\cite{he}, for two-electron systems,
$\tilde{v}_{\rm x}([\rho_{\rm HF}];{\bf r})=v_{\rm x}([\rho_{\rm HF}];{\bf r})$
and, consequently,  $v_{\rm c}^{\rm A}({\bf r})=v_{\rm c}^{\rm B}({\bf r})$.
Further, it was empirically found that the difference between
$v_{\rm c}({\bf r})$ and $v_{\rm c}^{\rm A,B}({\bf r})$
is small on the scale of $v_{\rm c}({\bf r})$.

\begin{figure}[htbp]
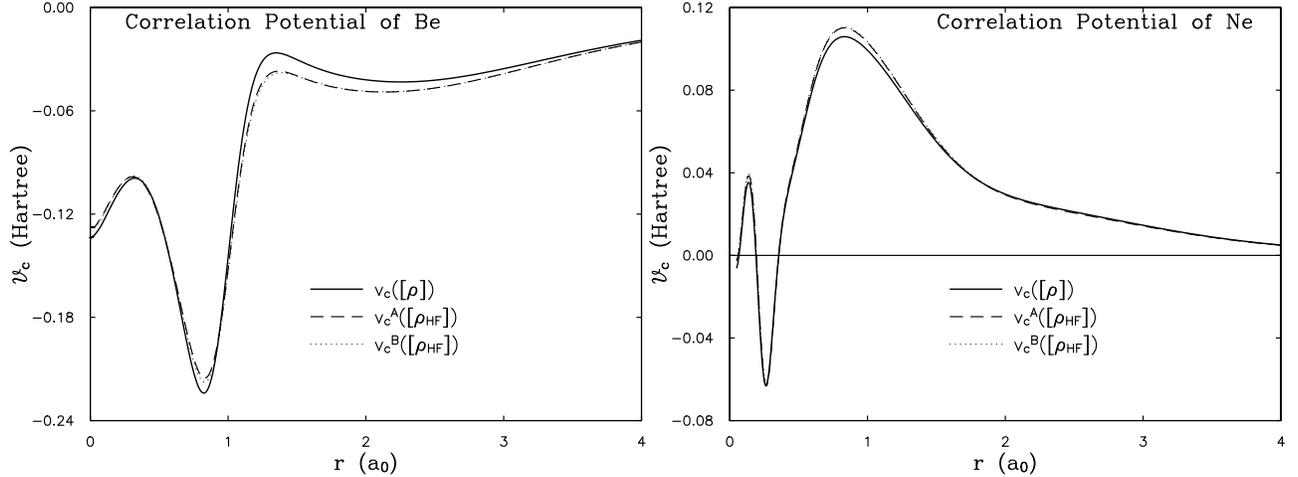

\centerline{\epsfxsize=8.5 cm \epsfbox{pl.be.vc}
            \epsfxsize=8.5 cm \epsfbox{pl.ne.vc}}
\caption[]{Comparison of the correlation potentials of Be and Ne.
$v_{\rm c}$ is the correct correlation potential from Eq.~\ref{vcorr},
 $v_{\rm c}^{\rm A}({\bf r})$ and $v_{\rm c}^{\rm B}({\bf r})$ are the
approximate correlation potentials constructed from the HF density
using Eq.~\ref{vcorrapprox}.
$v_{\rm c}^{\rm A}({\bf r})$ and $v_{\rm c}^{\rm B}({\bf r})$ are nearly
indistinguishable on the scale of these plots.
For Ne, there is some uncertainty in the potentials for $r<0.4\;a_0$.
}
\label{be.vc}
\end{figure}

Here, we find that even for the many-electron atoms Be and Ne, the
difference between $v_{\rm x}([\rho];{\bf r})$, 
$v_{\rm x}([\rho_{\rm HF}];{\bf r})$ and 
$\tilde{v}_{\rm x}([\rho_{\rm HF}];{\bf r})$
are almost not visible on the scale of $v_{\rm x}([\rho];{\bf r})$.
As shown in Fig.~\ref{be.vc}, the difference between
$v_{\rm c}^{\rm A}({\bf r})$ and $v_{\rm c}^{\rm B}({\bf r})$
is just barely visible even on the more expanded scale of $v_{\rm c}({\bf r})$,
This agreement is expected since the HF and the OEP densities are very close 
to each other and, for the OEP density, the agreement would be perfect.
On the other hand, the difference between either
$v_{\rm c}({\bf r})$ and $v_{\rm c}^{\rm A}({\bf r})$ or
$v_{\rm c}({\bf r})$ and $v_{\rm c}^{\rm B}({\bf r})$ is visible.
For both atoms, the exact and approximate potentials are clearly different,
although the shapes are very similar.
The similarity of the exact and the approximate potentials justifies the use of
the approximate scheme used in earlier work.


\section*{Acknowledgements}
We benefited from useful discussions with Mel Levy, Andreas G\"orling and 
Eberhard Gross.
The calculations were performed on the IBM SP2 computer at the Cornell Theory
Center.
This work is supported by the Office of Naval Research and NATO (grant number
CRG 940594).
X.G. acknowledges financial support from FNRS-Belgium and the
European Union (Human Capital and Mobility Program contract CHRX-CT940462).

\appendix
\section{Closed-shell case}
\label{a1}

In Sec.~\ref{s1}, we presented the theory of separation of the 
exchange-correlation potential into the exchange and correlation components 
for spin-unpolarized systems. In the present section, we restrict ourselves
to the special case of a closed shell atom.
For a closed shell atom, the self-consistent solutions of the Kohn-Sham 
equations (Eqs.~\ref{KS} and~\ref{rho}) can be factorized as the product 
of radial and angular components:
\begin{eqnarray}
\psi_i({\bf r})=R_i(r)Y_{l_i m_i}(\hat{\bf r})=\frac{\phi_i(r)}{r}
Y_{l_i m_i}(\hat{\bf r}),\label{orb}
\end{eqnarray}
where $l_i$ is the angular momentum quantum number.
Using this expression for the orbitals, the density (Eq.~\ref{rho}) can be 
rewritten as a sum over the occupied shells:
\begin{eqnarray}
\rho(r)=\frac{1}{4\pi r^2}\sum_{i=1}^{N_s} f_i\phi_i^2(r),
\end{eqnarray}
where $N_s$ is the number of occupied shells and $f_i$ is the occupation number 
of the $i$-th shell, $f_i=2(2l_i+1)$.

Following the derivation by Slater~\cite{Slater}, we rewrite the exchange 
energy (Eq.~\ref{enx1}) as
\begin{eqnarray}
E_{\rm x}\left[\rho\right]=-\sum_{i,j}^{N_s}\sqrt{(2l_i+1)(2l_j+1)}
\sum_{k=\left|l_i-l_j\right|}^{l_i+l_j}
c^k(l_i,0;l_j,0)\,G^k(n_i,l_i;n_j,l_j),\label{enxr}
\end{eqnarray}
where the coefficients $c^k$ incorporate the integrals over $\theta$ and 
are tabulated in Ref.~\cite{Slater} and $G^k$ is given by
\begin{eqnarray}
G^k(n_i,l_i;n_j,l_j)=\int {\rm d}r_1\int {\rm d}r_2\;
\phi_i(r_1)\,\phi_j(r_2)\,\phi_j(r_1)\,\phi_i(r_2)\,\frac{r_<^k}{r_>^{k+1}},
\end{eqnarray}
with $r_<=\min\{r_1,r_2\}$ and $r_>=\max\{r_1,r_2\}$.

The functional derivative of the exchange energy with respect to the effective 
Kohn-Sham potential (Eqs.~\ref{oep1} and~\ref{oep2}) can here be obtained
taking into account that the density depends only on the radial components 
of the Kohn-Sham orbitals. Eq.~\ref{oep1} is therefore equivalent to
\begin{eqnarray}
\frac{\delta E_{\rm x}\left[\rho\right]}{\delta v_s(r)}=
2\int_0^\infty{\rm d}r' \,v_{\rm x}(r')\;
\sum_{i=1}^{N_s} f_i\,\phi_i(r')\frac{\delta \phi_i(r')}{\delta v_s(r)},
\label{oep1p}
\end{eqnarray}
while Eq.~\ref{oep2} reduces to
\begin{eqnarray}
\frac{\delta E_{\rm x}\left[\rho\right]}{\delta v_s(r)}=
\int_0^\infty{\rm d}r'\,\sum_{i=1}^{N_s}
\frac{\delta E_{\rm x}\left[\rho\right]}
{\delta \phi_i(r')}\,\frac{\delta \phi_i(r')}{\delta v_s(r)}.
\label{oep2p}
\end{eqnarray}
Eqs.~\ref{oep1p} and~\ref{oep2p} can be combined to give the 
following integral equation for the exchange potential:
\begin{eqnarray}
\int_0^\infty{\rm d}r'\;v_{\rm x}(r')\,{\cal K}(r',r)={\cal Q}(r).
\end{eqnarray}

The functional derivative of the exchange energy (Eq.~\ref{enxr}) with respect 
to the radial orbital is given by
\begin{eqnarray}
\frac{\delta E_{\rm x}\left[\rho\right]}{\delta\phi_i(r)}
=-2\,f_i\sum_{j=1}^{N_s}\sqrt{\frac{2l_j+1}{2l_i+1}}
\sum_{k=\left|l_i-l_j\right|}^{l_i+l_j} c^k(l_i,0;l_j,0)\,
\phi_j(r)\int_0^\infty dr_2\;\phi_i(r_2)\phi_j(r_2)\frac{r_<^k}{r_>^{k+1}}.
\end{eqnarray}

The functional derivative of the radial orbital, $\delta \phi_i(r)/\delta 
v_s(r')$, is expressed in terms of the Green's function $G_i(r,r')$ as
\begin{eqnarray}
\frac{\delta \phi_i(r)}{\delta v_s(r')}=-G_i(r,r')\phi_i(r'),
\end{eqnarray}
where $G_i(r,r')$ satisfies the following differential equation:
\begin{eqnarray}
\left\{-\frac{1}{2}\frac{{\rm d}^2}{{\rm d}r^2}+\frac{l_i(l_i+1)}
{2 r^2}+v_s(r)-\epsilon_i\right\}\,G_i(r,r')=\delta(r-r')
-\phi_i(r)\phi_i(r').\label{g.rad}
\end{eqnarray}
This equation can also be derived by starting from Eq.~\ref{g.gen} in 
polar coordinates and projecting out the radial component. It can be easily 
checked that $G_i$ has the following expression:
\begin{eqnarray}
G_i(r,r')=\sum_{j\neq i: l_j=l_i}
\frac{\phi_j(r)\phi_j(r')}{\epsilon_j-\epsilon_i},
\end{eqnarray}
where the sum is over all the orbitals, except the $i$-th one, with angular 
momentum quantum number $l_i$.
In solving the differential equation for $G_i$, 
we set $r\neq r'$, divide by $\phi_i(r')$ and determine $\chi_{\rm out}
(r)$ and $\chi_{\rm in}(r)$ as solutions of outward ($r<r'$) and inward 
($r>r'$) integration. $\phi(r)$ is a homogenous solution of Eq.~\ref{g.rad} 
and can be added to $\chi_{\rm out}(r)$ and $\chi_{\rm in}(r)$
as $\alpha_{\rm out}\phi(r)$ and $\alpha_{\rm in}\phi(r)$ respectively. 
The difference $\alpha_{\rm out}-\alpha_{\rm in}$ is determined by imposing
continuity on $G_i$ and the sum $\alpha_{\rm out}+\alpha_{\rm in}$ by 
requiring that
\begin{eqnarray}
\int_0^\infty {\rm d}r\; \phi_i(r)\frac{\delta \phi_i(r)}{\delta v_s(r')}=0,
\end{eqnarray}
which follows from the normalization of $\phi_i(r)$.
Finally, we obtain 
\begin{eqnarray}
\frac{\delta \phi_i(r)}{\delta v_s(r')}=
\tilde{G}_i(r,r')-\phi_i(r)\int_0^\infty{\rm d}r''\;\phi_i(r'')
\tilde{G}_i(r'',r'),
\end{eqnarray}
where $\tilde{G}_i(r,r')$ is
\begin{eqnarray}
\tilde{G}_i(r,r')&=&
\theta(r-r')\left\{\chi_{\rm in}(r)
\phi_i(r')^2\, -\, \frac{1}{2}\left[\chi_{\rm in}(r')-\chi_{\rm out}(r')\right]
\phi_i(r)\phi_i(r')\right\}\nonumber\\
&+&\theta(r'-r)\left\{\chi_{\rm out}(r)
\phi_i(r')^2+\frac{1}{2}\left[\chi_{\rm in}(r')-\chi_{\rm out}(r')\right]
\phi_i(r)\phi_i(r')\right\}.
\end{eqnarray}

\end{document}